\documentclass[prb,nofootinbib,twocolumn,superscriptaddress]{revtex4-1}
\usepackage{graphics}
\usepackage{graphicx}
\usepackage{amsmath}
\usepackage{amsfonts}
\usepackage{amsthm}
\usepackage{amssymb}
\usepackage{amsbsy}
\usepackage{wasysym}
\usepackage{bm}
\usepackage{mathrsfs}
\usepackage{color}

\newcommand{\beq}{\begin{equation}}
\newcommand{\eeq}{\end{equation}}
\newcommand{\bea}{\begin{eqnarray}}
\newcommand{\eea}{\end{eqnarray}}

\def\ehat{{\hat en}}

\def\ehat{\hat{\mathbf{e}}}

\def\sel{\text{spinful electrons} }
\def\sel{{e\sigma}}

\def\ket#1{{\left|#1\right\rangle}}
\def\bra#1{{\left\langle #1 \right|}}
\def\braket#1{{\left\langle #1|#1\right\rangle}}
\def\br{\mathbf{r}}

\def\bq{\mathbf{q}}

\def\bR{\mathbf{R}}

\def\dg{^\dagger}

\def\be{\begin{eqnarray}}
\def\ee{\end{eqnarray}}

\newcommand{\smallsection}[1]{\subsection*{#1}}

\newcommand{\optionaltext}[1]{#1}  
\newcommand{\nonarxivversioncitation}[1]{}  

\def\onlinecite{\cite}

\begin{document}

\title{Featureless and non-fractionalized Mott insulators on the honeycomb lattice at $1/2$ site filling}

\author{Itamar Kimchi}
\affiliation{Department of Physics, University of California, Berkeley, CA 94720, USA}
\author{S. A. Parameswaran}
\affiliation{Department of Physics, University of California, Berkeley, CA 94720, USA}
\author{Ari M. Turner}
\affiliation{Institute for Theoretical Physics, University of Amsterdam, Valckenierstraat 65, 1018 XE Amsterdam, The Netherlands}
\affiliation{Department of Physics and Astronomy, The Johns Hopkins University, Baltimore, MD 21218}
\author{Fa Wang}
\affiliation{International Center for Quantum Materials and School of Physics, Peking University, Beijing 100871, China}
\author{Ashvin Vishwanath}
\affiliation{Department of Physics, University of California, Berkeley, CA 94720, USA}
\affiliation{Materials Science Division, Lawrence Berkeley National Laboratories, Berkeley, CA 94720}

\begin{abstract}
Within the Landau paradigm, phases of matter are distinguished by
spontaneous symmetry breaking. Implicit here is the assumption that a completely
symmetric state exists: a paramagnet. At zero temperature such quantum featureless insulators may be forbidden, triggering either conventional order or topological order with fractionalized excitations.
Such is the case for interacting particles when the particle number per \textit{unit cell}, $f$, is not an integer. But, can lattice symmetries forbid featureless insulators even at integer $f$?
An especially relevant case is the honeycomb (graphene) lattice --- where free spinless fermions at $f=1$ (the two sites per unit cell mean $f=1$ is half filling per site) are always metallic. 
Here we present wave functions for bosons, and a related spin-singlet wave function for spinful electrons,  
on the $f=1$ honeycomb, and demonstrate via quantum to classical mappings that they do form featureless Mott insulators. The construction generalizes to \textit{symmorphic} lattices at integer $f$ in any dimension. 
Our results explicitly demonstrate that in this case, despite the absence of a non-interacting insulator at the same filling, lack of order at zero temperature does not imply fractionalization. 
\end{abstract}

\maketitle

In describing phases of matter within the Landau paradigm, two phases 
connected by a continuous transition are related by 
the spontaneous breaking of an underlying symmetry.  Implicit in the paradigm is the assumption that a completely symmetric parent state ---  in the context of spin systems, a paramagnet --- exists.
In classical systems, this is just the high temperature phase. However, when
discussing quantum phases at zero temperature, a paramagnetic state could
be forbidden. For example, according to the Lieb-Schultz-Mattis  \cite{Lieb:1961p1}
theorem, for the spin-$1/2$ Heisenberg antiferromagnet in one dimension, no
completely symmetric quantum paramagnet exists. A spin disordered state
with exponentially decaying spin correlations will necessarily break
lattice symmetries.   An extension of this theorem \cite{Hastings:2004p1} applies to two
dimensions. In the  square lattice spin-$1/2$ Heisenberg model, which has a
half-odd-integer spin per unit cell, the spin disordered phase is not a
trivial paramagnet. If it does not break lattice symmetries then it must
be a quantum spin liquid, which has a hidden form of order called
topological order. The latter leads to fractionalized excitations with
novel  statistics, and is distinct from the notion of a featureless
paramagnet. The absence of symmetry breaking can then be taken as
indirect confirmation of the quantum spin liquid, which is often
used as a diagnostic both in numerics and in experiments \cite{BalentsSLRev}.  Furthermore, in
these systems where a trivial paramagnet is forbidden, quantum phase
transitions often lie outside the Landau-Ginzburg-Wilson paradigm \cite{SenthilScience}.
Therefore it is important to understand exactly when such trivial
paramagnets are disallowed.

These considerations can be readily translated to boson systems in
a periodic lattice \cite{Fisher:1989p1,Lindner2011}, as realized by ultracold atomic gases in optical
lattices \cite{Greiner:2002p1}. We assume a homogenous system, and define the {\em unit cell filling $f$} as the number of bosons per unit cell. 
When $f$ is not an integer, the
ground state must break some symmetry, e.g. by forming a superfluid or
enlarging the unit cell, or realize a topologically ordered state \cite{Oshikawa:2000p1,Hastings2005}.
(Consider that free fermions at noninteger $f$ possess partially filled bands and so must be metallic.) 
On a simple lattice with one site per unit cell,  a
Mott insulating state can appear at integer filling of bosons \cite{Fisher:1989p1}. This is the
bosonic analog of the trivial paramagnet. In a simple caricature of this
state, exact for some point within the phase, 
each site is occupied by a fixed integer number of bosons. 
Clearly no such classical picture exists 
when the number of bosons per site is not an integer, such as 
for $f=1$ on lattices with more than one site per unit cell. 

Let us consider the obstacles to featureless insulators in tight-binding models with multiple sites per unit cell and at $f=1$, i.e. fractional site filling.  
Clearly, a uniform integer filling of each site is impossible. 
For some lattices, one can identify one or more sites within each unit cell that are collectively invariant under the action of all point group symmetries and are not shared with any other cell.  
A ``molecular orbital'' product state, in which a boson is superposed symmetrically across that set of (one or more) sites for each unit cell, breaks no symmetries and is manifestly insulating since each boson is localized in a finite region. 
However, such situations  are not generic and for many lattices such as the kagome and honeycomb, no symmetric molecular orbital exists. For instance, identifying such a molecular orbital for each unit cell 
 on the honeycomb requires a choice between three orbitals related by $2\pi/3$ rotations;
the product state resulting from such a choice has correlations breaking $2\pi/3$ rotation symmetry. Another possibility for a symmetric insulator is a topologically ordered phase where emergent excitations
carry a fraction of the boson charge, giving one fractionalized quasiparticle per site.
But such fractionalization spoils featurelessness. 
The kagome and honeycomb lattices 
\footnote{Throughout this paper, we consider only {\it tight-binding} models where particles are restricted to occupy sites on the given lattice. This is crucial to distinguish between the triangular, honeycomb and kagome lattices, which share the same space group symmetries.} at $f=1$
admit no simple recipes for featureless insulators.

In a recent publication  \cite{pitv-arxiv} we demonstrated a procedure for generating bosonic featureless insulators as analogues of fermionic band insulators. 
The construction relies on using the free fermion band insulator to define a set of orthogonal, exponentially localized Wannier orbitals that respect lattice symmetries, and occupying each one with a boson. 
It applies to the kagome at $f=1$ (1/3 boson per site), for which there exists a corresponding band insulator.  On the honeycomb lattice at $f=1$ (one half particle per site), free spinless fermions can \emph{never} form a band insulator;  Dirac cones in graphene are symmetry-protected, as we discuss below. 
This rigorous lack of a $f=1$ fermionic band insulator then suggests that a featureless insulating boson state might also turn out to be  prohibited on the $f=1$ honeycomb.

In this manuscript we demonstrate that a novel bosonic wave function, which we term the Voronoi permanent, can be constructed on the honeycomb at $f=1$ and shown to be a featureless insulator, without either symmetry breaking or topological order.  
This  insulating state 
has inherent strong site-occupancy fluctuations and admits no corresponding free-particle analogue.
We then present evidence that taking the hard core boson limit, in which we forbid multiple occupancy of a site, preserves the featureless insulator. The resulting state is a quantum paramagnet of $S=1/2$ with U(1) $S^z$ rotation symmetry. Moreover, it then also provides an SU(2) symmetric featureless insulator wave function of spinful electrons at half filling on the honeycomb lattice. 

Interest in the $f=1$ honeycomb lattice has been further spurred by a recent numerical study \cite{Meng2010} of 
its spinful electron Hubbard model. 
A phase at intermediate coupling appeared to be insulating and spin gapped but without symmetry breaking of any kind, and was proposed to be a spin liquid with topological order. 
Regardless of the ultimate fate of this particular Hubbard model, similar results for a generic honeycomb Hubbard model at half filling could now have an alternate explanation in terms of our spinful electron wave function, without necessitating fractionalized excitations in the bulk. 
Thus explicit signatures of fractionalization, such as ground state degeneracy or topological entanglement entropy \cite{GroverEETODet, BalentsTEE, Schollwoeck}, are necessary for distinguishing between fractionalized spin liquids and featureless insulating states related to the one presented here.

Defining the wave function is simple:  it is a symmetrized product state 
(permanent\footnote{Permanents are the fully-symmetrized analogues of determinants})
over the smallest symmetric orbital associated with each unit cell, which for the honeycomb is simply an equal amplitude superposition of the six sites around a hexagon. This hexagon orbital is a \textit{Voronoi cell} in that it involves the sites closest to the hexagon center, so that it is fully symmetric and overlaps only with neighboring hexagons. We thus call the wave function 
a \textit{Voronoi permanent}\footnote{We note that permanent wave functions have been considered for magnetization plateaus on the anisotropic triangular lattice by T. Tay and O. Motrunich, Phys. Rev. B {\bf 81}, 165116 (2010).}. Since sites are shared between hexagons, the wave function may develop various orders in the thermodynamic limit, necessitating explicit computation to determine its properties.

\section{Free fermions and honeycomb lattice symmetries}
Consider spinless  {\it fermions} on the honeycomb lattice. The nearest-neighbor tight-binding model yields the famous graphene spectrum, of Dirac cones at the two Brillouin zone corners ($\pm \mathbf{K}$). At half-filling with one fermion per unit cell, the Fermi energy is at the Dirac points. In fact the band touching at the $\pm \mathbf{K}$ points holds for any tight binding Hamiltonian on this lattice; in order to gap out a Dirac point, a lattice symmetry (reflection, threefold rotation or inversion) must be explicitly or spontaneously broken. 
To see this, we study
 the irreducible representations of the `little group' of the Dirac points. At any momentum $\bq$ in the Brillouin zone, the little group $G_\bq$ is the subgroup of the space group that leaves $\bq$ invariant or translates it by a reciprocal lattice vector; the Bloch Hamiltonian $h_\bq$ at $\bq$ commutes with the little group generators and so we can classify energy bands using irreducible representations of $G_\bq$.

Consider the $\mathbf{K}$ point; our arguments apply equally well to $-\mathbf{K}$. 
The $\mathbf{K}$ point is left invariant by $2\pi/3$ rotation as well as mirror reflection, so the respective operators  $R_{2\pi/3}$ and $\sigma_2$ together generate the little group ($G_\mathbf{K}\cong D_{3h}$). 
Acting on the two sublattice components of the Bloch function at $\mathbf{K}$, the symmetry operations are represented by the $2 \times 2$ matrices
\be
R_{2\pi/3} = \left(\begin{array}{cc} e^{i2\pi/3}&0\\0&e^{-i2\pi/3}\end{array}\right),\,\,\,\, \sigma_2 = \left(\begin{array}{cc} 0&1\\1&0\end{array}\right) . \nonumber
\ee
These form a two-dimensional irreducible representation, and thus the band touching is protected by symmetry.  This argument carries through for spinful fermions with SU(2) spin rotation symmetry, in which case half-filling corresponds to two fermions per unit cell. Thus, there is no fermionic band insulator at half-filling on the honeycomb lattice that preserves all its symmetries.
This rigorous conclusion rules out the route to Bose insulators constructed as counterparts to fermionic band insulators using their Wannier orbitals \cite{pitv-arxiv}. But a featureless Mott insulator turns out to still be possible, even with no possible band insulator counterpart, as we describe below.

\section{Candidate Honeycomb Mott State}
As discussed earlier, our candidate $f=1$ wave function is a product state over the minimal orbitals respecting lattice symmetries, involving the six sites around a hexagon,
\be
\ket{\Psi_{\hexagon}} = \prod_{\bR}  B^\dagger_{\bR}\ket{0} 
 \, , \,\,\,\,\,\,
 B^\dagger_{\bR} \equiv \frac{1}{\sqrt{6}}\sum_{j \in \hexagon_\bR}{ b^{\dagger}_{j}} \label{eq:Bdef}.
\ee
Here $j$ labels a site, $\bR$ labels a unit cell i.e. site on the Bravais lattice,  $\hexagon_\bR$ denotes the sites on the hexagon of the unit cell $\bR$, and $\ket{0}$ is the boson vacuum.  In first quantized form, the wave function is
\be
\label{eq:firstquantization}
\Psi_{\hexagon}(\br_1,\br_2,\ldots,\br_N) = \text{perm}\left[\phi_{\bR_i}(\br_j)\right]
\ee
where $\phi_{\bR}(\br) = \bra{0}b_{\br} B^\dagger_\bR \ket{0}$
is nonzero on the six sites of hexagon $\bR$, and $\text{perm}$ refers to the permanent. 
Note that since hexagons are in one-to-one correspondence with unit cells, and each unit cell has two sites, this state has the requisite $1/2$ boson per site.  This construction assigns a fully symmetric single-particle orbital to each unit cell, naturally generalizing to \textit{symmorphic} lattices, for which all the symmetries may be realized at a single point, as we elaborate below. A simple example is the analogous hexagon state on the kagome lattice, which hosts $1/3$ boson per site.

\section{Characterizing $| \Psi _{\hexagon}\rangle$ by a Loop Model mapping}
Because hexagons on neighboring unit cells overlap, 
$| \Psi _{\hexagon}\rangle$
 is a highly entangled state with no classical analogue.
 In the thermodynamic limit this entanglement may turn into superfluidity, or the state could split into a sum of 
  states with different symmetry breaking orders, i.e. which belong to different superselection sectors and thus cannot be combined into a single state.  
  We need to explicitly compute properties of $\ket{\Psi_{\hexagon}}$ such as the boson Green's function 
  $ \langle b_{i}^\dagger b_{j} \rangle$ 
which describes the ability of the particles to propagate between lattice sites $i$ and $j$. In a Mott insulator this decays exponentially,
 while for a 2D superfluid at $T=0$ it exhibits long range order. Actually as we demonstrate below we can map the correlations in $\ket{\psi_{\hexagon}}$ to those of a classical finite temperature 2D model with short-range interactions. True long-range order is thus ruled out by the Mermin-Wagner theorem; correlations must decay at least algebraically, corresponding to a Kosterlitz-Thouless (KT) superfluid phase. The only other possible ordering is a discrete breaking of lattice symmetry, which can be diagnosed by studying
   the spatial structure of correlations.
As an unconstrained product state, $\ket{\psi_{\hexagon}}$ has no emergent gauge field to host topological order; 
below we give further evidence for lack of topological order by showing adiabatic continuity between $\ket{\psi_{\hexagon}}$ and a trivial atomic insulator. 

Relating the ground-state wave function  to the partition function of a classical statistical mechanical model has notable precedents. These include the Laughlin  fractional quantum Hall wave function  \cite{Laughlin:1983p1}, the Rokhsar-Kivelson wave function of dimer models  \cite{Rokhsar:1988p1,sondhiMoessner} and the AKLT spin wave function  \cite{Affleck87, affleck1988vbg,PhysRevLett.60.531,SidDanShivaji}. 
We now show that the normalization $\braket{\Psi_{\hexagon}}$, appearing in the denominator of any boson correlation function, is the partition function of a classical statistical mechanical loop model. 
Details are in the Supporting Information supplement.  
Commuting a $B_{\bR'}$ from the bra $\bra{\Psi_{\hexagon}}$ across all $B^\dagger_{\bR}$ in the ket  $\ket{\Psi_{\hexagon}}$ leads to a sum over products of commutators matching $B$ to $B^\dagger$,
\be
\braket{\Psi_{\hexagon}}  
=  \sum_{\sigma} \prod_{\bR} [B_{\bR}, B^\dagger_{\sigma(\bR)}]\label{eq:normalization2}
\ee
where $\sigma$ denotes a permutation of the sites on the triangular lattice.  Since only neighboring hexagons share sites, the commutator is
\be\label{eq:commutator}
[B_{\bR},B^\dagger_{\bR'}] = \delta_{\bR,\bR'} + m \delta_{\bR', \text{nn}(\bR)}
\ee
where $\text{nn}(\bR)$ denotes the nearest neighbors of $\bR$ on the Bravais lattice.
Here  $m=\frac{1}{3} \ (m=\frac{1}{6})$ for the honeycomb (kagome) lattice. In general, $m=p/q$ where $p$ is the number of sites shared by a pair of neighboring Voronoi cells (e.g. hexagons), each of which has $q$ sites. 

The commutator (\ref{eq:commutator}) 
restricts permutations in (\ref{eq:normalization2}) to those in which each site $\bR$ is matched  to either (i) itself, contributing a multiplicative factor of $1$ to the weight (which we  represent as an empty site),  or (ii)  a neighboring site, contributing a factor $m$ (which we represent as an arrow pointing from $B$ to $B^\dagger$). Since every site must be matched to exactly one other site, the arrows form closed loops which cannot intersect or touch.
Thus the Voronoi permanent $\ket{\Psi_{\hexagon}}$ defines a statistical mechanical model of closed, nonintersecting directed loops (which we take to include empty sites and dimers) on the triangular (Bravais) lattice, in which each link in a loop configuration multiplies its probability by a factor $m$:
\be
\braket{\Psi_{\hexagon}} = \sum_{\text{loop configs}} e^{- \bar{H}_\text{cl}} \equiv \sum_\text{loop configs} m^{L_\text{total}} \ .
\label{eq:partitionfunction}
\ee
The inclusion of empty sites and dimers, and the constraint that loops cannot intersect or touch, distinguish this particular loop model from more conventional ones studied.
We note that $\ket{\Psi_{\hexagon}}$ admits a secondary classical mapping based on coherent states of bosons \cite{SupplementaryMaterial} which was used, together with perturbation theory \cite{SupplementaryMaterial} in $m$, to confirm loop model results.

\begin{figure}[t]
\includegraphics[width=8.1 cm]{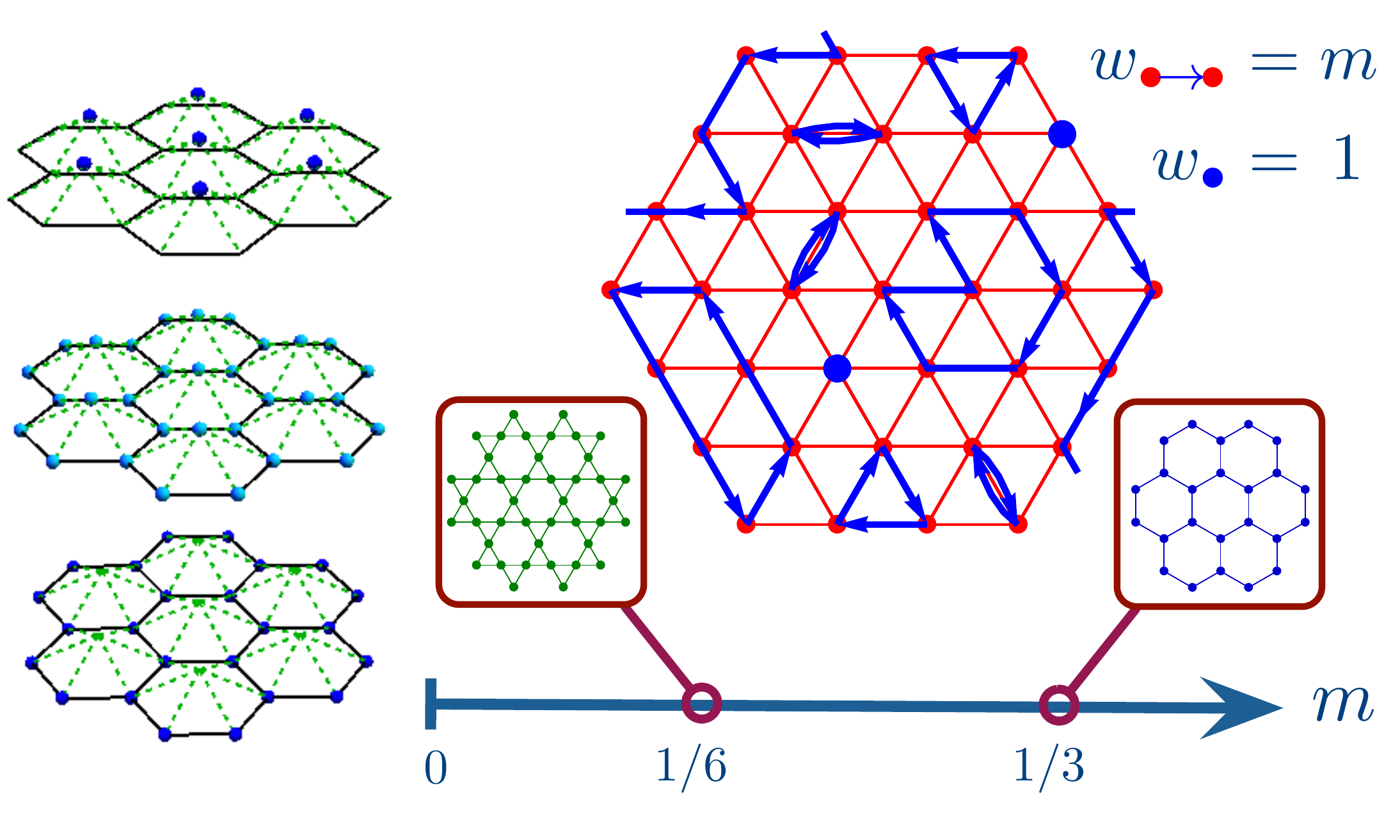}\caption{\label{fig:loops} {\bf Voronoi construction and loop mapping.}   (L) Decorated honeycomb and delocalization of bosons onto hexagons
; $m$ increases from $0$ to $1/3$ from top to bottom. (R) Sample loop configuration showing loops, dimers and empty sites which do not touch or intersect. Below we show the continuum of loop weights $m$, where $m=0$ is the atomic insulator and $m=\,1/6,\,1/3$  correspond to Voronoi permanents on the kagome and honeycomb.}
\end{figure}

This interpretation as a loop model unifies the Voronoi states on all lattices with the same underlying Bravais lattice. For the triangular Bravais lattice, the physical states on the kagome and honeycomb appear as specific points in the continuum of possible values for the loop weight $m$.
For an alternative continuum interpretation of $0<m<1/3$,
consider decorating the honeycomb lattice by adding a site (represented by $a^\dagger_\bR$) to the center of each hexagon, and modify the definition of $B^\dagger_\bR$, 
to $B^\dagger_\bR \rightarrow \cos\theta a^\dagger_\bR  + \sin\theta B^\dagger_\bR$. 
Since only  $b$ sites are shared by hexagons, computing the commutator now yields  (\ref{eq:commutator}) with  $m = \frac{1}{3}\sin^2\theta$.
For $\theta = 0$, we have an atomic insulator, since the bosons are restricted to the central site. As we increase $\theta$ we spread bosons across hexagons until we arrive at  the `honeycomb point' $\theta=\pi/2$  ($m=1/3$) where we can remove the empty central site.
  We have sketched this $m=0$ to $m=1/3$ interpolation of the Voronoi permanent in Fig. \ref{fig:loops}.
Thus, the mapping to the loop model lets us study whether correlation functions evolve smoothly, from the trivial $m=0$ atomic insulator with vanishing correlation length to the $m=1/3$ honeycomb state $\ket{\Psi_{\hexagon}}$.

We study the loop model
 using a modified classical Monte Carlo worm algorithm \cite{Svistunov2001}, an approach well-suited to this problem.
 First, the boson Green's function is mapped to a defect correlator of the loop model, exactly captured by an open worm. Second, the worm winding numbers directly yield the helicity modulus  (proportional to the superfluid density),
 a definitive diagnostic \cite{SupplementaryMaterial} 
of the KT superfluid phase, which is otherwise difficult to capture. This vanishes in the disordered phase and exhibits a universal jump of $\frac{2}{\pi}$ at the KT transition  \cite{Kosterlitz1977}.
 More complicated boson condensates (e.g. a `pair superfluid' where $\langle(b^\dagger_i)^2\rangle\neq 0$ but $\langle b^\dagger_i\rangle=0$) will also be captured in this approach.

\begin{figure}[t]
 \includegraphics[width=8.1 cm]{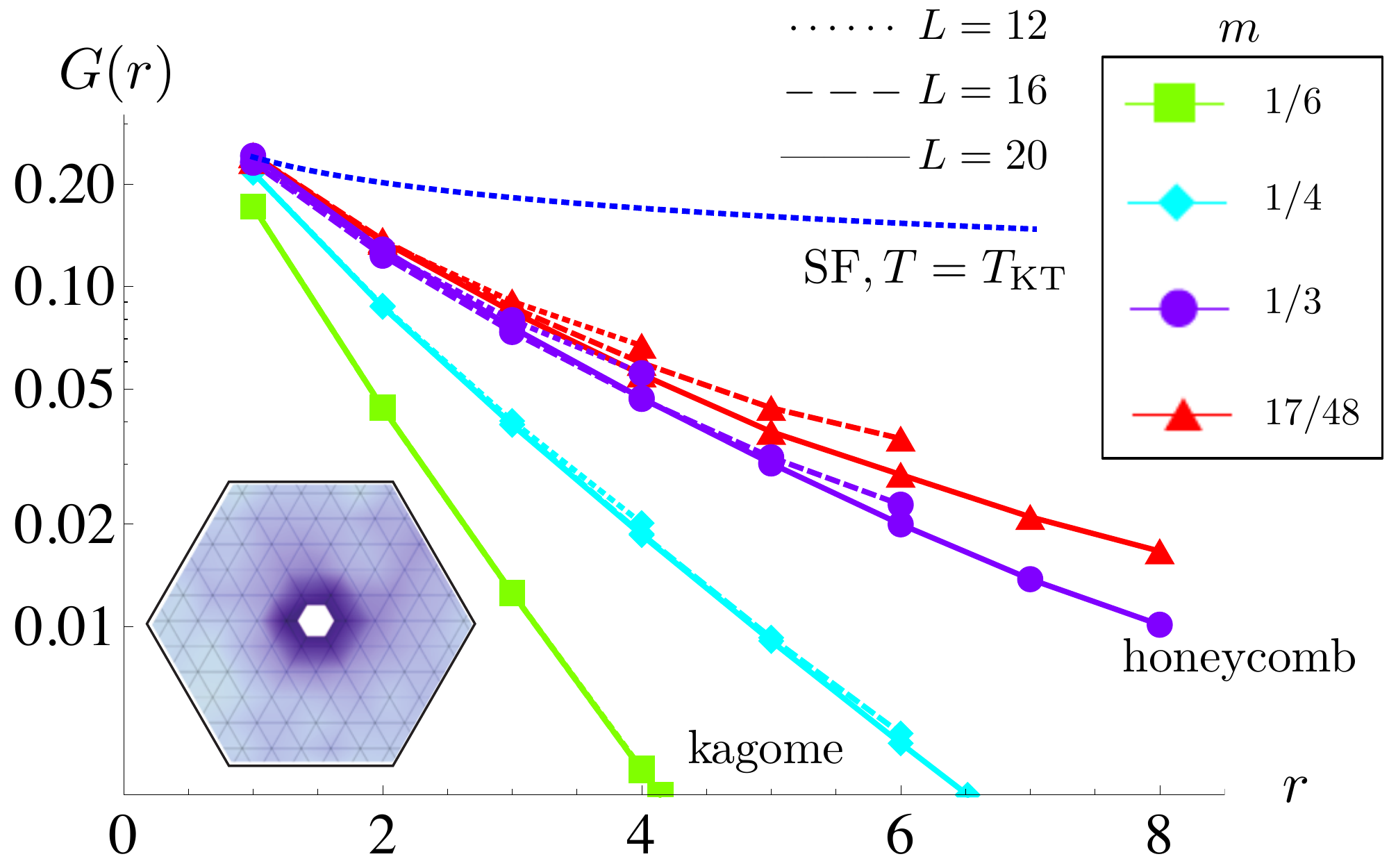}
\caption{{\bf \label{fig:CF}Loop model correlation function.}  Linear-log plot with distance measured along a triangular lattice basis vector. The fastest possible KT algebraic decay, $\sim r^{-1/4}$, is shown for comparison; $m=1/3$ (corresponding to the honeycomb lattice) displays exponential decay, indicating an insulating phase. Error bars are smaller than the line widths. 
Inset: Interpolated contour plot of loop model correlation function at $m=1/3$, with  white space in the center corresponding to the central peak. Correlations decay rapidly and are consistent with lattice symmetries.}
\label{fig:corrfn_SB}
\end{figure}

\begin{figure}[t]
\includegraphics[width=7.5 cm]{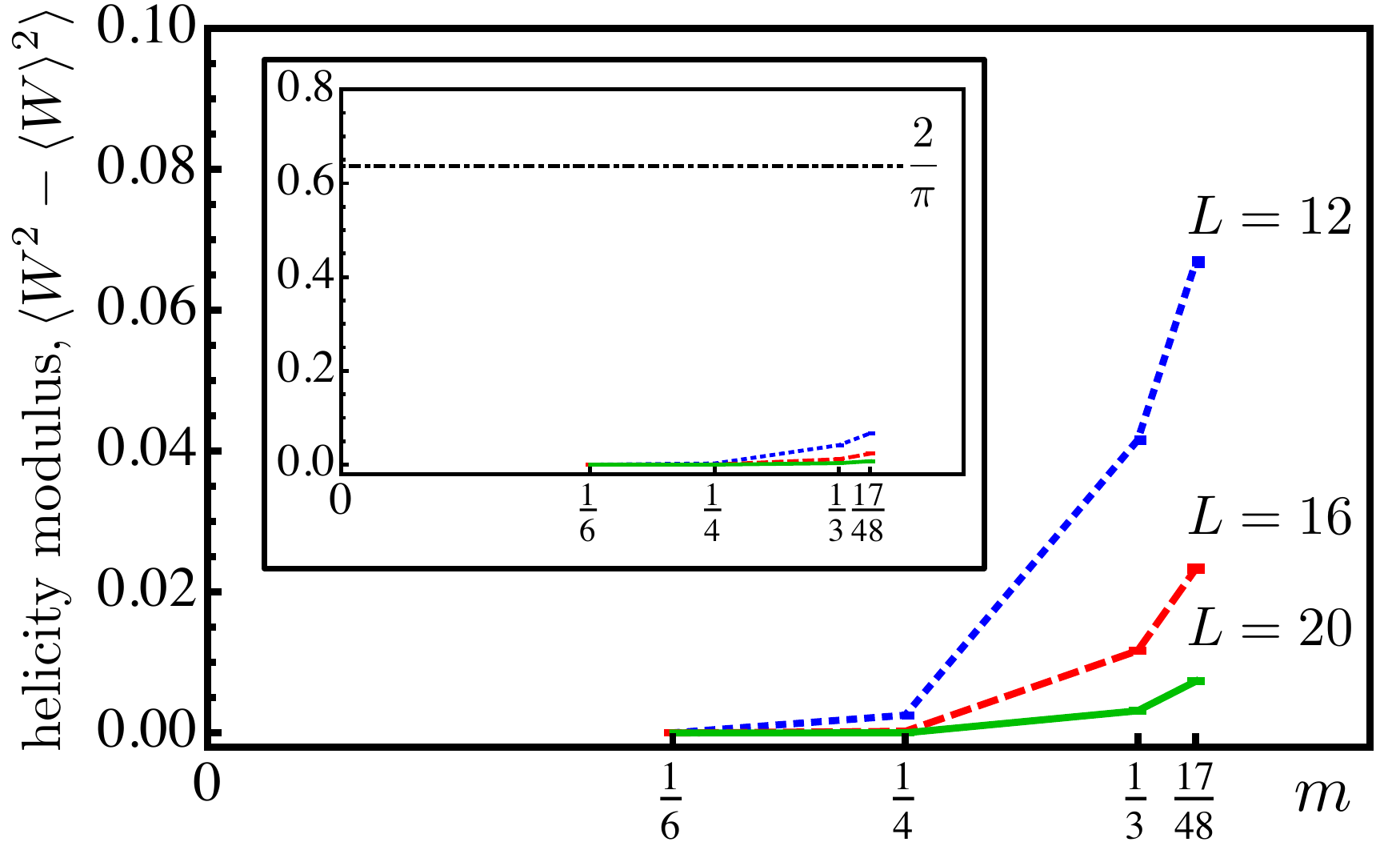}
\caption{{\bf \label{fig:winding} Helicity modulus.}   The helicity modulus (proportional to the superfluid density) tends to zero for increasing $L$, consistent with the expectation from KT finite-size scaling for the insulating phase.
Inset: Same figure, with axes zoomed out to show  that the helicity moduli for all $m$ values studied are much smaller than the universal  value of $2/\pi$ at the KT transition (the smallest superfluid density allowed in the KT phase). Error bars are smaller than the markers.}
\end{figure}

\smallsection{Numerical Results for $| \Psi _{\hexagon}\rangle$}   \ 
Our results are summarized in Figs. \ref{fig:corrfn_SB} and \ref{fig:winding}. 
We performed extensive worm algorithm Monte Carlo simulations for periodic $L\times L$ triangular lattice systems, with $L=12, 16, 20$, and averaged over $10^6$ Monte Carlo steps per site (MCS) in each case ($10^5$ MCS were sufficient for the kagome at low $m=1/6$).  
Computing the boson Green's function in $| \Psi _{\hexagon}\rangle$ we find it decays exponentially with a correlation length of  $\xi \sim 2.4$ lattice sites, as shown in Fig.~\ref{fig:corrfn_SB}.  The  correlation length evolves
  from $\xi=0$ at $m=0$ to $\xi\sim0.9$ lattice sites at the kagome ($m=1/6$) and remains a small fraction of the system size  for $m$ values beyond the honeycomb. 
  We also plot for comparison the most rapid possible algebraic decay in the KT phase, $G(r) \sim r^{-1/4}$, to emphasize that algebraic decay of single boson correlations is ruled out.
    As shown in Fig. \ref{fig:winding}, the superfluid densities  are all much less (see inset) than the universal jump value $2/\pi$ at the KT transition \cite{Kosterlitz1977}, indicating that the wave function remains in the insulating phase for the $m$ values studied. 
Finally, neither the loop model nor the coherent state simulations exhibit breaking of discrete lattice symmetries for the honeycomb lattice Voronoi permanent. 
We used short MC runs to avoid averaging out any symmetry-breaking by oversampling, and characterized lattice symmetry breaking \cite{SupplementaryMaterial}.  
 As a simple visual demonstration of lattice symmetry, a sample correlation function for the loop model on the triangular lattice at $m=1/3$ from an $L=12$, 500 MCS  short worm algorithm run is inset in Fig.~\ref{fig:corrfn_SB}.
We benchmarked results against coherent state simulations at $m=1/3,1/6$ and loop perturbation theory at small $m$ \cite{SupplementaryMaterial}.


\section{Projected wave function of hard core bosons}
The bosonic wave function $|\Psi_{\hexagon}\rangle$ may be interpreted as a spin wave function where
different boson occupations are different $S_z$ values, and only U(1) spin
rotation about the $z$ axis is a symmetry. The honeycomb lattice wave function
then represents a featureless magnetization plateau \cite{Gaulin2008} at $2/3$ of the
saturation value, with $S=3/2$ spins on the sites. A related $S=1/2$ wave function may be constructed by projecting out multiple boson occupation from $|\Psi_{\hexagon}\rangle$ (hardcore bosons limit). The resulting state $|P\Psi_{\hexagon}\rangle$ has hard core bosons at the same filling $f=1$, and can also be interpreted as a U(1)-symmetric spin-$1/2$ state.

Because $|\Psi_{\hexagon}\rangle$ is a Mott insulator and projection further suppresses particle number fluctuations (enhancing phase disorder), 
 $|P\Psi_{\hexagon}\rangle$ is unlikely to become a superfluid; we check this below. But projection may introduce 
 lattice symmetry breaking order such as a charge density wave (CDW) or valence bond solid (VBS). 
To check for these we study the particle number correlator 
$\langle \tilde{n}_i \tilde{n}_j \rangle$, where  $\tilde{n}_i = b\dg_i b_i -1/2$ gives density fluctuations above the mean.

\begin{figure}[t]
\includegraphics[width=8.1 cm]{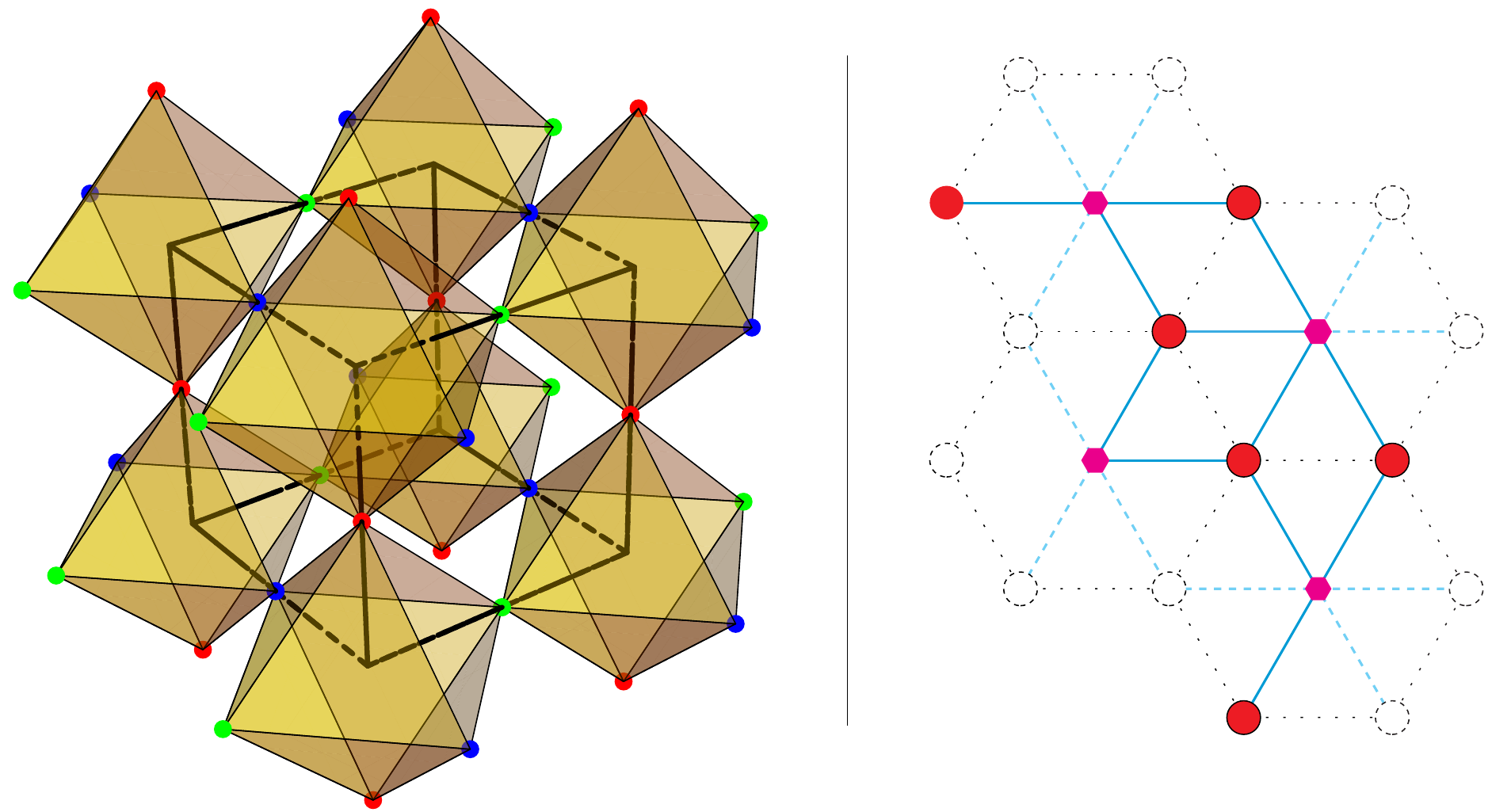}
\caption{{ \bf Left: Corner sharing octahedron lattice.}   A
symmorphic lattice with sites at vertices of corner-sharing octahedra. A Voronoi permanent wave function at $f=1$ can be constructed as a candidate insulator.
{\bf Right: Depleted dice lattice.} 
From the honeycomb lattice (dotted lines), construct the dice lattice by 
introducing bonds (solid and dashed lines) connecting the original honeycomb sites to additional new sites in hexagon centers (small magenta hexagons). Discard the original honeycomb bonds. We then deplete the dice lattice by removing unoccupied honeycomb sites (empty circles) and associated bonds (dashed light-blue). 
Classical dimers on the remaining graph (solid blue bonds) give the density correlator of the projected state. 
}
\label{fig:oct_and_dice} 
\end{figure}

This correlator is diagonal in the basis of boson occupancy configurations  $\{n_i \}$; 
the overlap $\langle \{n_i \} |P\Psi_{\hexagon}\rangle$, i.e. the number of terms within  $|P\Psi_{\hexagon}\rangle$ with boson numbers $\{n_i \}$, may be computed, in another classical mapping, by counting certain dimer configurations.  
 Consider the ``dice lattice'',   shown in Figure \ref{fig:oct_and_dice}. Given a boson occupancy configuration $\{n_i \}$ we deplete the dice lattice by removing unoccupied honeycomb sites and associated bonds.  On this depleted dice lattice, a dimer on a bond between hexagon center $R_i$ and honeycomb site $r_j$ selects the boson from the orbital of $R_i$ to occupy the site $r_j$. Since dimers cannot touch, no more than one boson may occupy a given site, as required by the projection. Counting dimer coverings gives the overlap. 
 
Because the depleted dice lattice is planer and bipartite, its dimer coverings may be counted efficiently by the Kasteleyn-Fisher-Temperley algorithm \cite{Kasteleyn1961,Fisher-Temperley,Kenyon2009}. Each bond $\ell$ is assigned a weight $w_\ell = \pm 1$ such that the product of weights around any closed loop obeys
\beq
\prod_{\ell \in \text{loop}} w_\ell =  (-1)^{1+\frac{1}{2}\sum_{\ell  \in \text{loop}}} .
\eeq 
Let $D$ be the adjacency matrix of the depleted dice lattice, modified by putting the appropriate sign $w_\ell$ on every bond. 
Then, for open boundary conditions, the overlap is 
\beq
\langle \{n_i \} |P\Psi_{\hexagon}\rangle = 6^{-\sum n_i} \left| \text{det D}  \right|
\eeq
where det is the determinant. 
Details including periodic boundary conditions are described in the Supporting Information supplement. 
 
\begin{figure}[t]
 \includegraphics[width=8.1 cm]{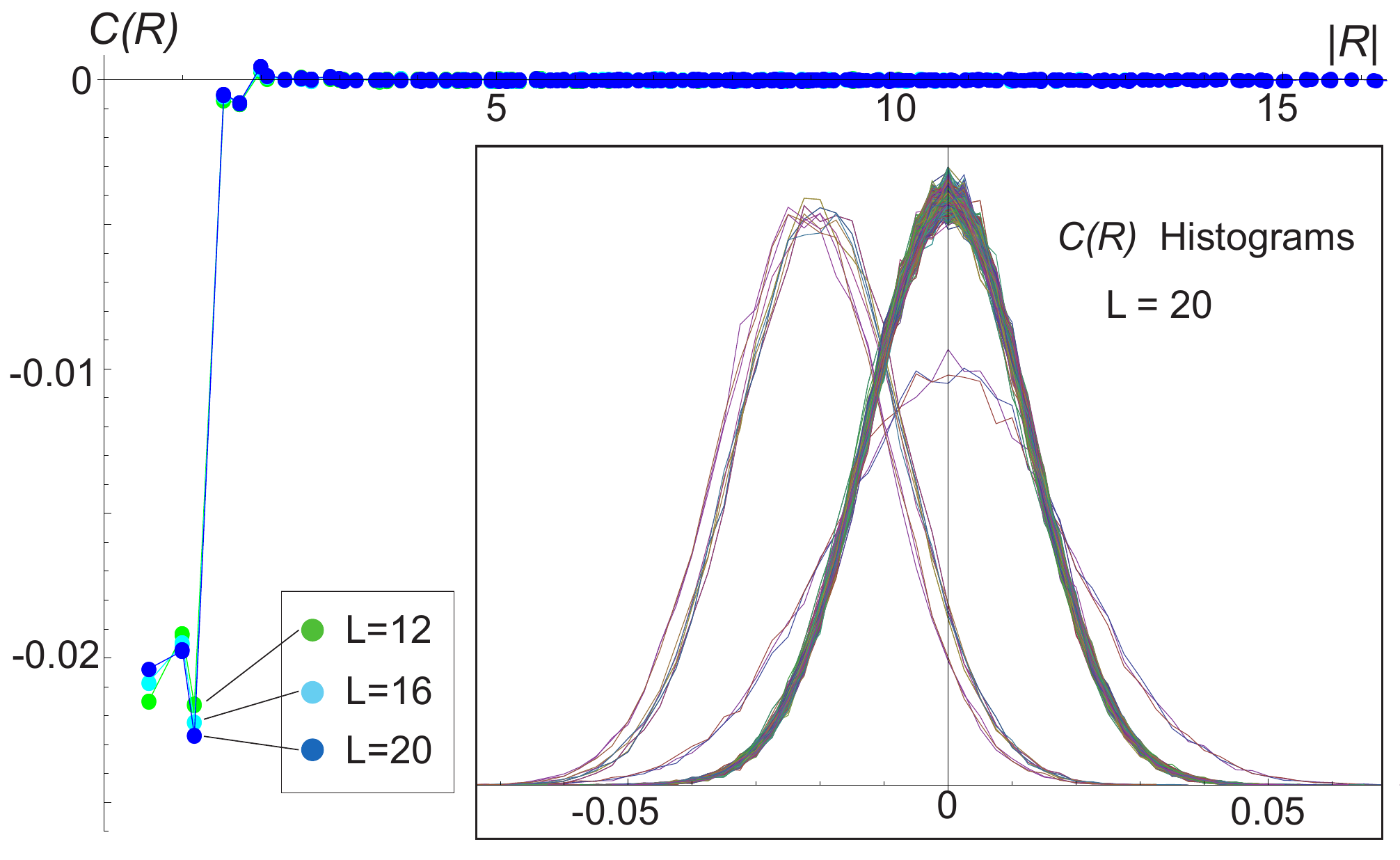}
\caption{{\bf \label{fig:projected-correlator} Density correlator in $|P\Psi_{\hexagon}\rangle$.} 
The correlation function $C(R)$ of particle number fluctuations in the projected state $|P\Psi_{\hexagon}\rangle$ versus distance $|R|$, exhibiting rapid decay beyond third neighbors.
 Error bars are smaller than the line widths.  (Inset) The histogram of correlations $C(R)$ for every pair separation $\vec{R}$ in the $L=20$ system. 
The histogram for each correlator is a single-peaked Gaussian, providing strong evidence for lack of lattice symmetry breaking.}
\end{figure}

We compute the determinant while sampling $\{n_i \}$ using determinantal Monte Carlo \cite{Kalos1977,Gros198953}. The results are shown in Figure~\ref{fig:projected-correlator}. We find that the density-density correlator is short ranged with vanishing amplitude beyond a few sites, and moreover that the histogram of correlation between any pair of sites is unambiguously single-peaked. 
This correlator decays algebraically \cite{SupplementaryMaterial} 
 (as $1/r^3$) for any superfluid, so
the decay shown in Fig.~\ref{fig:projected-correlator}, which does not appear algebraic, suggests the state is insulating. 
Moreover the rapidly decaying and single-valued behavior of the correlator rules out any ordering which would break symmetry in density correlations, such as charge density wave and valence bond solid. 

Finally, since we find that projection takes one spinless boson insulator state to another, both with short ranged correlations and no symmetry breaking, we expect that the entanglement remains short ranged as the bosons become hard core. Then no topological entanglement entropy or fractionalization could emerge upon projection.  We conclude that 
 $|P\Psi_{\hexagon}\rangle$, a hard core boson or U(1)-symmetric spin-$1/2$ state, is indeed gapped and featureless.


\smallsection{SU(2)-symmetric spinful electron featureless insulator}  \ 
The primary significance of $|P\Psi_{\hexagon}\rangle$ is that it immediately yields a wave function for spinful electrons. A hard core boson operator $b\dg_i$ has exactly the same algebraic structure as an on-site Cooper pair of electrons, $c\dg_{i,\uparrow} c\dg_{i,\downarrow}$. The boson state $|P\Psi_{\hexagon}\rangle$ thus enables the construction of a spinful electron ($e\sigma$) state, in a different Hilbert space, but with the same featureless characteristics,
\beq
\label{eq:electronstate}
|P\Psi_{\hexagon}\rangle  \Longleftrightarrow  | \Psi_\sel \rangle \equiv  \prod_R \sum_{j \in \hexagon_R }  c\dg_{j,\uparrow} c\dg_{j,\downarrow} |0\rangle \ .
\eeq
Here $j$ labels sites and $R$ unit cells; the spinful electrons are at half filling. This state respects all lattice as well as full SU(2) spin symmetries.  It is fully gapped, since even single electron excitations are gapped by the s-wave pairing.  The Cooper pairing $c\dg_{j,\uparrow} c\dg_{j,\downarrow}$ was already determined to be short ranged in studying the bosonic state, so this electron state is not a superconductor but an insulator.

\section{Discussion and outlook} 
We began this paper by posing a question: is a featureless and non-fractionalized (i.e., topologically
trivial) insulating phase  possible  for a system of bosons  on the
honeycomb lattice at a filling of  one boson per unit cell?  By explicitly
constructing a simple trial wave function (the Voronoi permanent) and numerically computing correlations within
it, we have argued that such a phase does indeed exist.

On the way, we have also constructed a Voronoi permanent wave function at filling one on the kagome lattice, without any order.   On the kagome lattice, a similar state can be constructed in a simpler way (see Ref.~\onlinecite{pitv-arxiv}), using Wannier orbitals from a band insulator, eliminating the need for explicit numerical computation. The hexagon orbitals can be viewed as a truncation of the Wannier orbitals of this kagome band insulator. 
But on the honeycomb lattice this simpler approach is impossible, since its free fermion band structure cannot be insulating
without breaking symmetry. 
A fermionic Slater determinant of Voronoi orbitals (hexagons) on the kagome lattice yields a kagome band insulator; but if we attempt to make a Slater determinant of Voronoi orbitals on the honeycomb lattice, the wave function vanishes.

The Voronoi wave functions obtained on both the kagome and the honeycomb are positive definite, and
could be ground states of an unfrustrated 
model, evading the sign problem and admitting efficient simulation by quantum Monte Carlo. 
However, we are currently unable to provide local Hamiltonians for which 
our wave functions are (unique) exact ground states.  
Analytically constructing such exact parent Hamiltonians \cite{Haldane1983,Kivelson1985} is a nontrivial exercise. Nevertheless, even in the absence of explicit parent Hamiltonians,  the wave functions $| \Psi _{\hexagon}\rangle$, $|P \Psi _{\hexagon}\rangle$ and $| \Psi _\sel\rangle$ with exponentially decaying correlations provide compelling evidence for an extended $f=1$ featureless insulator phase.   Indeed at fractional $f$ where Hastings' theorem\cite{Hastings:2004p1} prohibits such a phase for any local Hamiltonian, our construction also does not apply.
 Another promising avenue is to work with simple model Hamiltonians with judiciously chosen parameters 
and numerically search for featureless insulating phases corresponding to our wave functions.
Ultracold atoms have recently been confined
to kagome \cite{Jo:2012p1} and honeycomb \cite{Tarruell:2012p1} optical lattices, 
further motivating 
 numerical studies of
realistic Bose-Hubbard model Hamiltonians that may realize
such fractional site filling insulators.

As pointed out above, the construction of the honeycomb Voronoi state naturally generalizes to a construction applicable to all symmorphic lattices; the resulting states are candidate potential featureless insulators. Symmorphic lattices are defined by having a symmetry group which \textit{splits}: the point group and the translations are independently generated. This implies the existence of a symmetric orbital at each unit cell. 
That makes it possible to construct Voronoi states from symmetric orbitals that overlap only on neighboring unit cells, i.e. Voronoi orbitals. 
To test whether a given state is featureless, correlators and superfluid densities are again computable using a loop mapping. 
Many Voronoi states on lattices sharing the same underlying Bravais lattice map to the same classical loop model, at different effective temperatures $1/m$.
To study such Voronoi states with a given Bravais lattice, it is sufficient to simulate a single loop model.  

On non-symmorphic lattices it is impossible to choose a fully symmetric orbital at each unit cell and the
construction fails.   Examples of non-symmorphic lattices include the well
known pyrochlore and diamond lattices.
Recent work has shown that featureless insulators at $f=1$ are forbidden on non-symmorphic lattices \cite{nonsymmorphic}, suggesting that the Voronoi states exhaust the possibilities for featureless insulators at $f=1$.

In addition to the honeycomb, a related
2D lattice is the checkerboard lattice with two sites in the unit
cell, which may be viewed as a set of corner sharing tetrahedra in two
dimensions. Again the (spinless) fermonic bands must touch due to lattice
symmetries \cite{SunYaoKivelsonFradkin}, and the Voronoi permanent construction is required.  A similar 3D example is given by the cubic
perovskite lattice of corner-sharing octahedra (Fig~\ref{fig:oct_and_dice}) with three sites per unit cell.  
A possible 2D example with a band touching was considered in Ref.~\onlinecite{YaoKivelson}, but in this case it is trivial to construct a Bose insulator 
via the `molecular orbital' approach of filling non-overlapping sets of sites.

The Voronoi wave function can also be interpreted as a spin wave function where
different boson occupations are different $S_z$ values, and only U(1) spin
rotation about the $z$ axis is a symmetry. The honeycomb state
then represents a featureless magnetization plateau at $2/3$ of the
saturation value, with $S=3/2$ spins on the sites. The projected hard core boson state corresponds to $S=1/2$ with symmetry again reduced to U(1). 
A state with full SU(2) spin symmetry was also found as a paired spinful electron analogue of the hard core bosons, where spin-up and spin-down electrons on the same site pair into an SU(2) singlet.  This electron featureless insulator Eq.~\ref{eq:electronstate} has on average half an electron of each spin per site, meaning it may arise in a honeycomb Hubbard model at half filling of sites. 
Note however that a bipartite Hubbard model with purely on-site interactions, such as that studied in Ref.~\onlinecite{Meng2010}, has the larger symmetry group SO(4)$=$SU(2)$\times$SU(2)$/\mathbb{Z}_2$;
the paired electron state Eq.~\ref{eq:electronstate} only has the physical SU(2) spin symmetry so it is not a candidate featureless state for that model, though it can be the featureless ground state for more generic  interactions. 
In all these cases, the featureless insulators that we find explicitly demonstrate that for the relevant systems, a full gap with complete lack of symmetry breaking does not imply fractionalization.

Of the assortment of boson, spin and electron featureless insulator wave functions we have presented for the honeycomb lattice at $f=1$, there is a conspicuous absence: a spin wave function with $S=1/2$ and full SU(2) spin rotation symmetry.  The SU(2) symmetric electron state Eq.~\ref{eq:electronstate}  has either zero or two (paired) electrons on a site, so it vanishes upon Gutzwiller projection to single occupancy. 
Whether a different approach will yield an SU(2) symmetric $S=1/2$ honeycomb quantum paramagnet, or whether such a spin state is rigorously forbidden, remains an open question.

\begin{acknowledgments}
We thank Bryan Clark, Matthew Fisher, Tarun Grover, Olexei Motrunich, Shivaji Sondhi and Matthias Troyer for useful discussions, and Dan Stamper-Kurn and Dan Arovas for collaboration on related work. We also thank an anonymous referee for constructive comments. 
This research is supported in part by the National Science Foundation under Grants No. DGE 1106400 and NSF PHY11-25915 for the KITP Graduate Fellowship Program (I.K.),  the Simons Foundation (S.A.P.) and the Army Research Office with funding from the DARPA Optical Lattice Emulator program (A.V.).
\end{acknowledgments}



\begin{appendix}
\section*{SUPPORTING INFORMATION}

\section{Loop Model Mapping}
We now expand on mapping a Voronoi permanent state to the statistical mechanical loop model.
Here we express a site $i \equiv(\bR, \alpha)$ by its sublattice index $\alpha$ and unit cell $\bR$, a point on the Bravais lattice.
We are guided by computing the boson Green's function 
\begin{equation}
G_{\alpha,\alpha'}^{bos}(\bR,\bR')=
\frac{\bra{\Psi_{\hexagon}}  b_{\bR,\alpha}^{\dagger}b_{\bR',\alpha'}\ket{\Psi_{\hexagon}}}{\braket{\Psi_{\hexagon}}}
\label{eq:corrfn}
\end{equation}
within the Voronoi permanent
\begin{equation}
| \Psi_{\hexagon} \rangle = \prod_{\bR} \left( \sum_{\bR',\alpha} f_{\bR}(\bR',\alpha) b^{\dagger}_{\bR',\alpha} \right) |0\rangle .
\end{equation}

For the kagome lattice, $\alpha = 1,2,3$ where we choose a unit cell in which the three sites belonging to a unit cell lie in the directions of the primitive Bravais lattice vectors $\mathbf{a}_1, \mathbf{a}_2, \mathbf{a}_3$ respectively, with $\mathbf{a}_1 + \mathbf{a}_2 + \mathbf{a}_3 = 0$.
For the hexagon state we then have $f_{\bR}(\bR',\alpha)=\frac{1}{\sqrt{6}}\left(\delta_{\bR,\bR'}+\delta_{\bR,\bR'+\ehat_\alpha}\right)$ where the
final $\ehat_\alpha$ refers to the Bravais lattice vector pointing from one
hexagon to an adjacent hexagon which shares sublattice site $\alpha$ with it. For our choice of unit cell,  $\ehat_\alpha=\mathbf{a}_\alpha$.

For the honeycomb, $\alpha =1,2$ where the unit cell is taken to be the top,bottom sites respectively on a vertical bond. Choose Bravais vectors $\mathbf{a}_1=(1,0)$, $\mathbf{a}_2=(-1,\sqrt{3})/2$, $\mathbf{a}_3 =-\mathbf{a}_1-\mathbf{a}_2$.  For the hexagon state we then have
 $f_{\bR}(\bR',\alpha)=\frac{1}{\sqrt{6}}\left(\delta_{\bR,\bR'}+\delta_{\bR,\bR'-\mathbf{a}_1} +  \delta_{\bR,\bR'+\mathbf{a}_{\alpha+1}}\right)$.

Consider the normalization (i.e. denominator of (\ref{eq:corrfn}))
\be
\braket{\Psi_{\hexagon}}&=&\bra{0}\left(\prod_{\bR}B_{\bR}\right) \left(\prod_{\bR'}B_{\bR'}^{\dagger}\right)\ket{0}\ee
with  $B_{r}^{\dagger}=\sum_{\bR',\alpha}f_{\bR}(\bR', \alpha)b_{\bR',\alpha}^{\dagger}$.
A single pair $B_{\bR'}^{\dagger} ,  B_{\bR}$ gives the commutator 
 $\left[ B_{\bR},B_{\bR'}^{\dagger}  \right] = C[\bR,\bR'] $ 
 $= \sum_{\bR_1,\bR_2,\alpha_1,\alpha_2} f^*_{\bR}(\bR_1,\alpha_1) f_{\bR'}(\bR_2,\alpha_2)$.
In a Wick type decomposition, this becomes a sum over all bijective maps i.e. permutations $\sigma:\bR \rightarrow \bR'$, with each term in the sum being the product of commutators $\prod_\bR C[\bR,\sigma(\bR)]$. We are saved from computing this functional integral because the hexagons of two unit cells overlap only if they belong to the same or neighboring unit cells, $C[\bR,\bR']=\delta_{\bR,\bR'} + m \delta_{\bR',nn[\bR]}$, so maps $\sigma$ only appear in the sum if they take $\bR$ either to itself (weight 1) or to a neighbor (weight $m$).

An allowed map $\sigma$ can be pictured as a collection of arrows between neighboring sites on the Bravais  lattice, with each site $\bR$ either having no arrows (in case $\sigma$ maps it to itself) or exactly one arrow going out (to $\sigma(\bR)$) and one arrow coming in (from $\sigma^{-1}(\bR)$). Therefore $\braket{\Psi_{\hexagon}}$ is a sum over directed closed loop configurations on the Bravais lattice, with each configuration in the sum weighted by $m$ to the power of the total length of its loops. Loops may not touch or intersect with the exception of the zero area length two loop, which is permitted. Strictly speaking this is a  generalized vertex model 
(with $37$ states per vertex, $6^2$ one-in-one-out states and one for the case when the site is mapped to itself) and not a simple loop model, 
 though we may still study it by the worm algorithm by incorporating two bond-occupation flavors. 

Next, we turn to the numerator. Each $b$ operator knocks out a $B^\dagger$ operator with a coefficient
$f$, resulting in a sum over {\it defect correlators},
\begin{equation}
G_{\alpha,\alpha'}^{bos}(\bR,\bR')=\sum_{\bR_1,\bR_2}f^*_{\bR_1}(\bR,\alpha)f_{\bR_2}(\bR',\alpha')G^{worm}(\bR_1,\bR_2)
\end{equation}
where we named the defect correlator $G^{worm}$ because of its natural interpretation as a correlation function in the worm algorithm for $\bR_1 \neq \bR_2$,
\begin{equation}
G^{worm}(\bR_1,\bR_2)=\frac{\langle0|\left(\prod_{\bR\neq \bR_1}B_{\bR}\right)\left(\prod_{\bR'\neq \bR_2}B_{\bR'}^{\dagger}\right)|0\rangle}{\braket{\Psi_{\hexagon}}}.
\end{equation}
For the kagome, the boson correlator is given by the sum of four worm correlator values; for the honeycomb, the sum of nine. Thus, we can extract correlation functions on the honeycomb and kagome lattices explicitly from the worm algorithm defect correlator.

\section{Loop Perturbation Theory}
Within first-order perturbation theory in $m$, loops are costly and the sum in the numerator will involve only configurations where the defects are connected by the shortest line segment between them.  To this order, restricting for a moment to a 1D lattice, we find that defect configurations are all multiplied by an extra factor of $m^{|R-R'|}$. This naive estimate holds when there is only one shortest line connecting the defects, as is the case when $R-R'$ is a multiple of a triangle lattice vector $R_1$ or $R_2$. However, consider the case where $R-R' = n(R_1-R_2)$. Then there are many shortest line segments between the defects, identical to the number of shortest paths between opposite corners of an $n$ by $n$ square lattice. There are ${2n}\choose{n}$ such paths, asymptotically going as $4^{n}$, each carrying weight $m^{2n}$. Thus we find that the asymptotic behavior of the correlation function, depending on its direction, lies between
\[  m^{\ell} \leq \langle b_0^\dagger b_{\ell} \rangle  \leq (2 m)^{\ell}  \]
within first order perturbation theory in $m \sim 1/T$. This suggests a transition only occurs at a large $m\approx 1/2$. Equality with the lower bound is achieved for any $\ell$ for sites separated by a multiple of a Bravais lattice basis vector. This perturbation theory result provided a further benchmark of this loop model worm algorithm.

\section{Worm algorithm computations}
The directed loop configurations with and without an open chain may be studied simultaneously using a Monte Carlo `worm' algorithm \cite{Svistunov2001}.
The basic idea of the worm algorithm is to simultaneously gather statistics on the correlation function and the normalization in (\ref{eq:corrfn}) by working directly in the loop representation. 
The `worm' is an open loop configuration; by allowing a worm to shrink or grow by a random process, until it closes of its own accord, while preserving detailed balance -- implemented by a local-update Metropolis algorithm --  one gains statistics on both open and closed configurations, which contribute respectively to the numerator and denominator of (\ref{eq:corrfn}). The canonical example usage of a worm algorithm is to study the XY model in the loop-current (dual) representation.

Within the worm algorithm simulations, each system was initialized in the infinite temperature `$m=0$' configuration with no loops, and equilibrated over 50 000 MCS before recording averages. Error estimates were obtained by computing the standard error of the data from $20$ independent runs, a conservative approach.  

Unlike the usual XY case, our loop model has strong interactions in that the loops are forbidden
to touch or intersect, necessitating some modifications to the algorithm. Here, we merely note that the twin complications of loop self-avoidance and the triangular lattice geometry presents unique challenges. At large   $m$ when long loops are favorable, the worm can occasionally get `stuck' in a configuration (for instance, a spiral) for which most proposed updates will be rejected. The worm would then fail to close and the algorithm fail to converge. This issue necessitated long runs ($\gtrsim 10^6$ MCS) for the honeycomb, though for smaller $m$ we found $10^5$ MCS is sufficient. For the honeycomb ($m=1/3$) there was a single run at the largest system size ($L=20$) that did not converge; we discarded it and a similar $m=17/48$ run when taking averages. We note that the honeycomb data for $L=12,16$ are free of such convergence issues.

\optionaltext{
\section{Helicity Modulus in Terms of Winding Numbers}
The relation between worm winding numbers and the helicity modulus may be seen as follows. Take the loop model on a torus and thread a flux $\theta$ through one of its handles, say 
the cycle associated with the periodic boundary condition $\vec{r}\sim \vec{r}+ L \hat{x}$ 
in the $\hat{x}$ direction. This is equivalent to a uniform vector potential $A\hat{x}$ permeating the system with magnitude $A=\theta / L $.
The Boltzmann factor for the worm to grow a step $\vec{\delta R}$ now appears multiplied by the phase $\exp(i A\hat{x}\cdot\vec{\delta R})$. This phase factor cancels out for any closed loop, {\it unless it threads the torus}, crossing the periodic boundary conditions with a nonzero \textit{winding number} $W_x \equiv \int \hat{x}\cdot\vec{\delta R} /L$. The winding number is integer-valued for closed loops. Let $W_x$ stand for the sum of all $\hat{x}$ winding numbers in a given closed loop configuration. Then
 the contribution of flux to the free energy,  $\Delta F[\theta] = F[\theta] - F[0]$ is given by
\begin{equation}
\exp\left(-\beta \Delta F[\theta]\right) = \langle \exp \left(i \theta W_x \right)
\end{equation}
yielding the helicity modulus \be
Y\equiv d^2 F/ d\theta^2 |_{\theta=0} = \beta^{-1} \left( \langle W^2 \rangle - \langle W \rangle^2 \right).\ee The superfluid inverse temperature $\beta$ which appears in these expressions is set to $1$ in our model.  Winding number fluctuations thus identify the superfluid.
}

\section{Bosonic Coherent States}
Another way to map the wave function $\ket{\Psi_{\hexagon}}$ is to work in the basis of bosonic coherent states. 
\optionaltext{
Let us consider the bosonic correlation function,
\be
G^{bos}(0,j) = \langle b^\dagger_0 b_j\rangle = \delta_{j0} + \frac{\bra{\Psi_{\hexagon}} b_j b^\dagger_0 \ket{{\Psi_{\hexagon}} }}{\braket{{\Psi_{\hexagon}} }}
\ee
}
We 
\optionaltext{use the resolution of identity for bosonic coherent states,
\be
\mathbf{1} = \int \left[dz_id\bar{z_i}\right] e^{-\sum_{i} |z_i|^2} \ket{\{z_i\}}\bra{\{z_i\}}
\ee
where $\ket{\{z_i\}} = \otimes_i \ket{z_i}$ and $\left[dz_id\bar{z_i}\right] =\prod_i \frac{d\text{Re}z_i d\text{Im}z_i}{\pi}.$ Inserting this above, and using the property of the coherent states that $b_i \ket{\{z_i\}} = z_i \ket{\{z_i\}}$ and $\left\langle0|\{z_i\}\right\rangle =1$ we 
}find
\be
\langle b^\dagger_0 b_j\rangle &=&\delta_{j0} + \frac{ \int \left[dz_id\bar{z_i}\right] e^{-\sum_{i} |z_i|^2}\prod_{\bR}|\mathcal{B}_{\bR}(z)|^2   \left({z}_j \bar{z}_0\right)   }{ \int \left[dz_id\bar{z_i}\right] e^{-\sum_{i} |z_i|^2}\prod_{\bR}|\mathcal{B}_{\bR}(z)|^2    }\nonumber\\
\ee
where $\mathcal{B}_{\bR}(z) =  \sum_{i} f_{\bR}(i) z_i$.
We see that it suffices to consider the correlation function $\langle {z}_\bR \bar{z}_0\rangle$ of the statistical mechanical model described by a complex number at each site, with the the manifestly positive semi-definite Boltzmann weight
$\mathcal{P}(\{z_i\}) = \prod_{\bR} |\mathcal{B}_{\bR}(z)|^2 e^{-\sum_i |z_i|^2}$.

The model's $U(1)$ symmetry admits an algebraically correlated KT phase, better captured by the loop model. Nevertheless, simulating the coherent bosons on system sizes up to $L=12$, using a simple local-update Metropolis algorithm, allowed us to benchmark worm results.

\section{Discrete Symmetry Breaking}
In addition to showing there is no transition into a superfluid, we must show there is no discrete (Ising) transition of discrete symmetry breaking. Performing a point group symmetry operation $U$ on the lattice transforms the site and sublattice parameters of the boson Green's function, yielding the `Ising' order parameter
\be
\mathcal{I}^{(U)}_{\alpha\beta}(\bR) \equiv G^{bos}_{\alpha\beta}(\bR)  - UG^{bos}_{\alpha\beta}(\bR)U^{-1} .
\ee
We computed the rms average of $\mathcal{I}^{(U)}_{\alpha\beta}(\bR)$ for the point group generators (rotations and reflections) using short Markov chains of 100 -- 2000 MCS, ensuring the chains do not sample multiple symmetry broken states while still collecting sufficient statistics.
To check for symmetry breaking we primarily used simulations of the wave functions in the coherent state representation, since fluctuations in the associated Markov chains exhibit shorter correlations than those in the worm algorithm. 
We found $\mathcal{I}^{(U)}_{\alpha\beta}(\bR)$ vanishes asymptotically within our error estimates.

\optionaltext{
\section{Additional Numerical Data}
In Fig.~\ref{fig:CFloglog} we show a log-log plot of the worm algorithm correlation function to a power law for the largest system size, $L=20$ (same data as in Fig.~\ref{fig:CF}) with the fastest KT algebraic decay shown for comparison on a log-log plot. It is clear that a power-law fit is inconsistent with the data as this would appear as a straight nearly-horizontal line on this plot.

\begin{figure}[tb]
\includegraphics[width=7.1 cm]{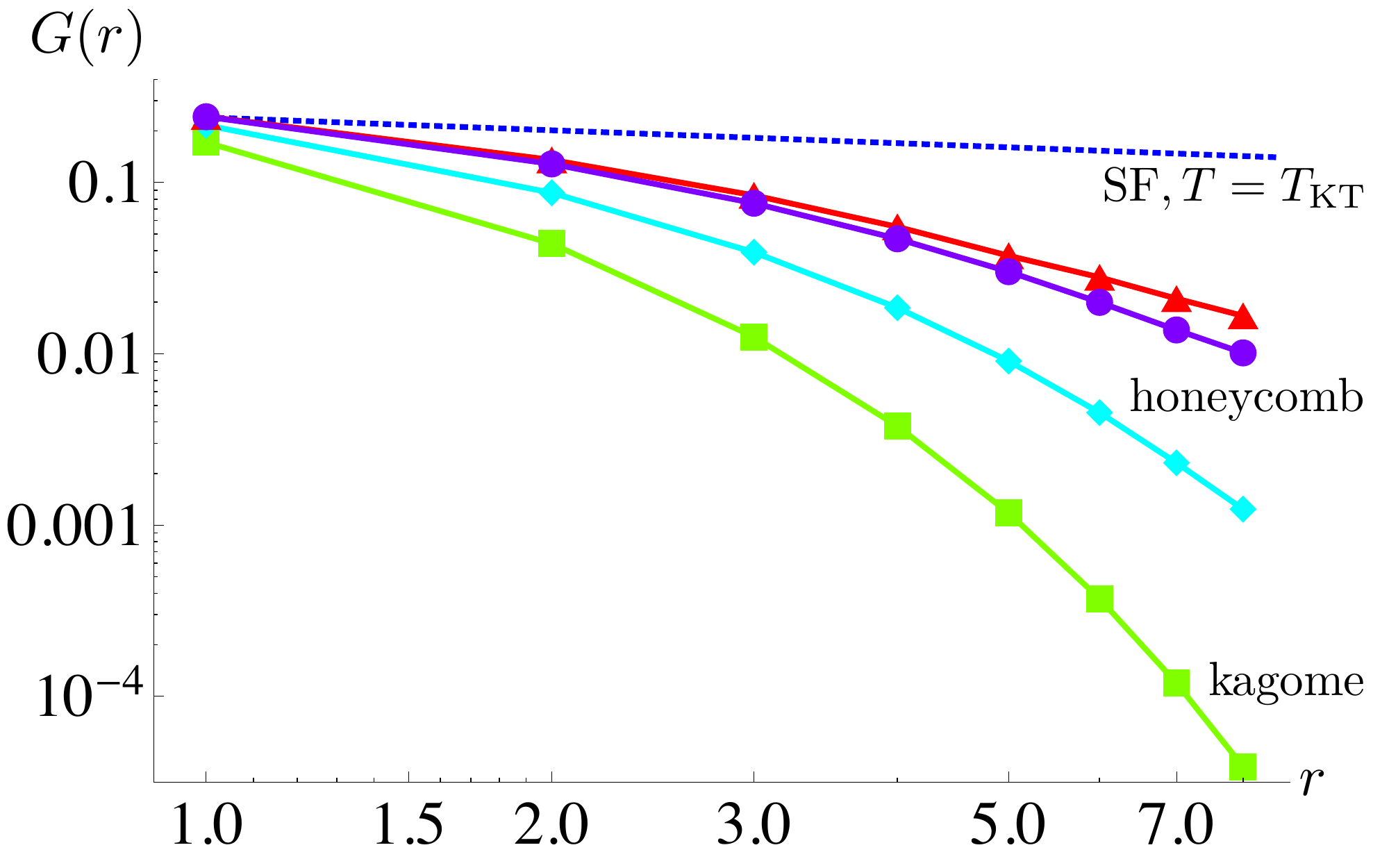}
\caption{{\bf \label{fig:CFloglog}Loop model correlation function.}  Log-log plot as a function of distance along a basis vector, demonstrating power-law fit is impossible. Error bars smaller than linewidths.}
\end{figure}
}

\section{Variational Monte Carlo with Kasteleyn Matrix Computation for the Projected State}
We now describe the Monte Carlo algorithm used for 
evaluating the density correlation functions of the Gutzwiller projected
boson wave function. In some aspects this algorithm is similar 
to the traditional determinantal Monte Carlo algorithm for Gutzwiller projected 
fermion wave functions \cite{Kalos1977,Gros198953}. 

\optionaltext{
The density correlator we compute is
 \be
C(i,j)=
\frac{
\left\langle P\Psi_{\hexagon} \left|
\left(b\dg_i b_i -\frac{1}{2}\right)
\left(b\dg_j b_j -\frac{1}{2}\right)
\right| P\Psi_{\hexagon} \right\rangle }
{\langle P\Psi_{\hexagon}|P\Psi_{\hexagon}\rangle} .
\label{eq:densitycorr}
\ee 
Denote the boson number on site $j$ as
$n_j^{\vphantom{\dagger}}=b_j^\dagger b_j^{\vphantom{\dagger}}$. 
The (un-normalized) projected boson wave function is formally
$|P\Psi_{\hexagon}\rangle=\prod_j (\delta_{n_j,1}+\delta_{n_j,0})|\Psi_{\hexagon}\rangle$, 
where $\delta$ is the Kronecker-$\delta$ symbol.
The expectation value of a diagonal operator $O$, {e.g.}
the density correlation function, is
\begin{equation}
\frac{ \langle P\Psi_{\hexagon}|O|P\Psi_{\hexagon}\rangle }
{\langle P\Psi_{\hexagon}|P\Psi_{\hexagon}\rangle}
=\sum_{\{n\}}
\frac{ |\langle\{n\}|P\Psi_{\hexagon}\rangle|^2 }
{\langle P\Psi_{\hexagon}|P\Psi_{\hexagon}\rangle}
\cdot \langle \{n\} |O|\{n\}\rangle,
\end{equation}
where $|\{n\}\rangle$ is a boson number basis state, 
and the sum is over all possible boson number configurations.
The first factor 
$|\langle\{n\}|P\Psi_{\hexagon}\rangle|^2/\langle P\Psi_{\hexagon}|P\Psi_{\hexagon}\rangle$, 
namely the weight of the boson number configuration $\{n\}$,
is then simulated by a Markov chain.
To this point the algorithm is identical to previous
determinantal Monte Carlo methods \cite{Kalos1977,Gros198953}. 
The new ingredient is the polynomial time algorithm for evaluating the 
overlap $\langle\{n\}|P\Psi_{\hexagon}\rangle$.
}

The overlap $\langle\{n\}|P\Psi_{\hexagon}\rangle$ with the number basis diagonalizing the density correlator is either zero (for $\{n\}$ with multiple site occupancy) or, just like the case for $\Psi_{\hexagon}$, is the permanent
\begin{equation}
\langle\{n\}|P\Psi_{\hexagon}\rangle=
\Psi_{\hexagon}(\boldsymbol{r}_1,\dots,\boldsymbol{r}_N)
={\rm perm}[\phi_{\boldsymbol{R}_i}(\boldsymbol{r}_j)],
\end{equation}
where $\boldsymbol{r}$ are boson positions, $N$ is the number of 
unit cells (also the number of bosons), 
and $\phi$ is given after Equation~\ref{eq:firstquantization}. 
\optionaltext{
$\phi$ is proportional to the adjacency matrix between 
hexagon center $\boldsymbol{R}$ and honeycomb sites $\boldsymbol{r}$,
\begin{equation}
\phi_{\boldsymbol{R}_i}(\boldsymbol{r}_j)=
\left\{
\begin{array}{ll}
\sqrt{1/6},&{\rm if\ }\boldsymbol{r}_j\in \hexagon_{\boldsymbol{R}_i}\\
0,&{\rm otherwise}
\end{array}
\right.
\end{equation}
For the projected wave function, the occupied sites $\boldsymbol{r}$
are all distinct. 
}
As described in the main text, the overlap is then proportional to 
(with a factor $6^{-N/2}$) the number of dimer coverings of a
depleted dice lattice\optionaltext{ in which all unoccupied honeycomb sites are removed}. 
\optionaltext{
 The bonds in the depleted dice lattice are between hexagon centers $R_i$ to occupied honeycomb sites $r_j$.  A dimer on a bond $R_i--r_j$ selects the boson from orbital $R_i$ to occupy the honeycomb site $r_j$.
}

The depleted dice lattice is a bipartite planar graph. 
Its dimer counting problem can be solved by 
a polynomial time algorithm \cite{Kasteleyn1961,Fisher-Temperley}
by converting the permanent to a determinant. 
With periodic boundary condition 
the permanent equals a linear combination of four determinants \cite{Kasteleyn1961},
\bea
&& 
{\rm perm}[\phi_{\boldsymbol{R}_i}(\boldsymbol{r}_j)] = 
 \\
&&
\frac{1}{2}
\left[ 
 {\rm det}(D_{+-})+{\rm det}(D_{--})  
+  
 {\rm det}(D_{-+})-{\rm det}(D_{++}) 
\right] . \nonumber
\eea
The matrices $D$ are defined by $(D_{p_x p_y})_{ij}=w_{i j} \phi_{\boldsymbol{R}_i}(\boldsymbol{r}_j)$
where $w_{i j} = \pm 1$ is the Kasteleyn weighting, 
and the first and second subscripts $p_x,p_y=+,-$ label the ``boundary condition''
along the horizontal and vertical directions respectively \cite{Kasteleyn1961}.
\optionaltext{

The key piece in the Kasteleyn-Temperley-Fisher algorithm \cite{Kasteleyn1961,Fisher-Temperley} is a weighting of the lattice by an orientation or equivalently a $\pm$ sign on each bond.   To construct this ``Kasterleyn orientation'' on the depleted dice lattice, 
we start from the full dice lattice with the
Kasterleyn orientation depicted in Fig.~\ref{fig:FullyFrustrated}.
We then remove the empty sites in pairs. 
When removing a pair of sites, 
we draw a line connecting them 
and cutting through an array of dice lattice bonds,
the orientation of these bonds should then be flipped. 
This line is the analogue of the string connecting two visons
in the Ising gauge theory \cite{Senthil-Fisher}.
An example of this procedure for the $++$ boundary condition 
is given in Fig.~\ref{fig:depleted-dice}.
Note that this ``vison string'' should  be chosen to not cross the boundary, 
otherwise it will insert a vison in a noncontractible loop of the torus
and change the $+/-$ boundary conditions.

\begin{figure}
\includegraphics[width=7.1 cm]{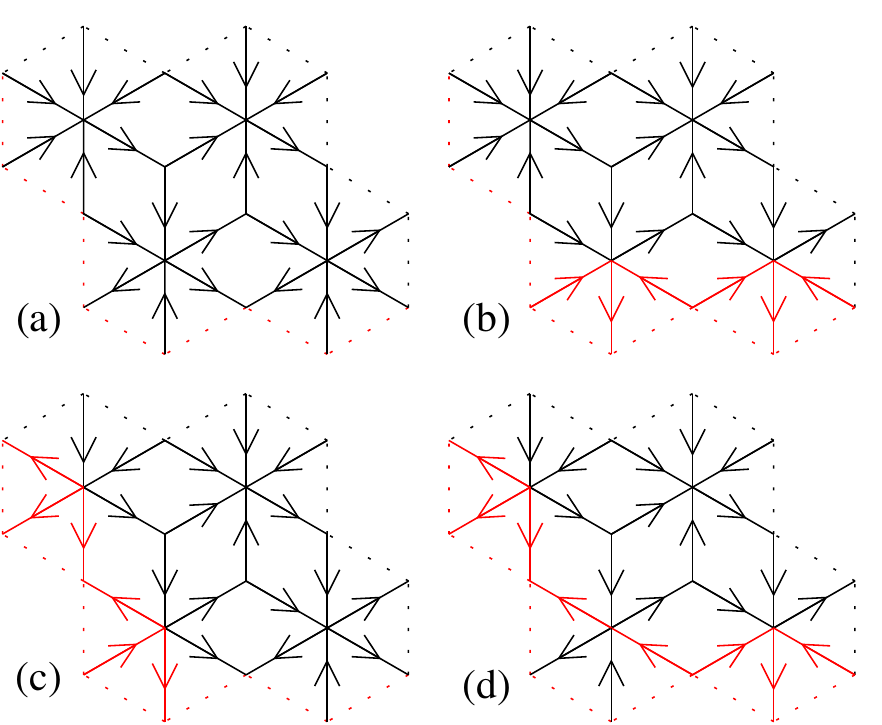}
\caption{{\bf Kasteleyn orientations with periodic boundaries. }
(Color online)
Kasteleyn orientation on the $2\times 2$ dice lattice with 
different boundary conditions,
(a) $++$, (b) $+-$, (c) $-+$, (d) $--$. 
Arrow from a hexavalent site $i$ to a trivalent site $j$ means
$D(i,j)=+\phi_{\boldsymbol{R}_i}(\boldsymbol{r}_j)$,
and the reverse arrow direction indicates the opposite sign.
Red dotted lines on the bottom and left boundary should
be identified with those on the top and right boundary.
Red bonds in (b)(c)(d) are different from (a) because
they cross either horizontal or vertical boundary.
Kasteleyn orientations of $++$ boundary condition 
for larger lattices are just the repetition of (a).
} \label{fig:FullyFrustrated}
\end{figure}

\begin{figure}
\includegraphics[width=7.1 cm]{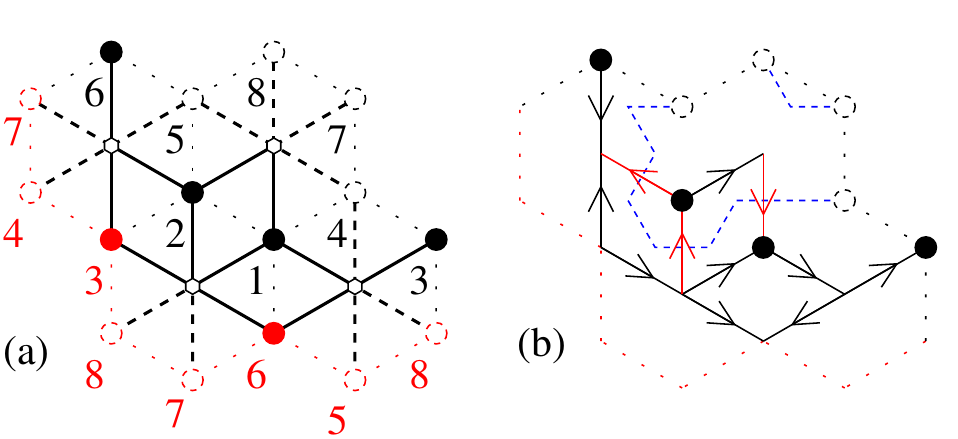}
\caption{{\bf Depleted dice lattice and Kasteleyn orientation. }
(Color online)
(a)
An example of the depleted dice lattice built from 
a $2\times 2$ honeycomb lattice with periodic boundary condition. 
$1,\dots,8$ label the eight honeycomb lattice sites. 
Dashed open circles($4,5,7,8$) are empty honeycomb sites.
Solid circles($1,2,3,6$) indicate boson occupied honeycomb sites.
Small open hexagons are hexavalent dice lattice sites 
(centers of hexagons in the honeycomb lattice).
Dotted lines are honeycomb lattice bonds.
Dashed lines are dice lattice bonds removed with empty honeycomb sites.
Solid lines are the remaining dice lattice bonds
forming the depleted dice lattice.
Red bonds and sites on the bottom and left boundary
should be identified with those on the top and right boundary.
(b)
The Kasteleyn orientation on this depleted dice lattice
with $++$ boundary condition.
Dashed blue lines are the ``vison strings'' connecting removed sites.
Red bonds are those different from 
Fig.~\ref{fig:FullyFrustrated}(a).
}
\label{fig:depleted-dice}
\end{figure}

}
The Monte Carlo moves are attempts 
to move a randomly selected boson to a random neighboring honeycomb site. 
\optionaltext{
The Metropolis rule is used to decide whether this move can be accepted.
Note that only one column of the matrices $\phi$ and $D$ will be changed by 
this move. 
Therefore the ratio of the weights before and after the move can be 
efficiently computed as in previous fermion Monte Carlo studies \cite{Kalos1977,Gros198953}.
This move is equivalent to moving an empty site, 
and the vison string attached to it should be modified accordingly. 
}
Note that if a boson is moved across the horizontal(vertical) 
boundary, the vertical(horizontal) boundary condition label $+/-$ should be 
flipped. 

Here we define a ``Monte Carlo sweep'' as 
$N$  attempted boson moves, where $N$ is the number of bosons.
For each Monte Carlo calculation, 
we start from a random boson state with nonzero weight,
 equilibrate the system by $10^4$ Monte Carlo sweeps, 
and then perform $2^{20} \approx 10^6$ measurements and Monte Carlo sweeps. 
Using a binning analysis, we extract autocorrelation times for each correlator $C(R)$, and find a maximum autocorrelation time of at most 1.7 sweeps, requiring no correction to the reported error bars, which are vanishingly small.

The density correlator in Figure \ref{fig:projected-correlator} is plotted against the magnitude of the distance $|r_i-r_j|$ between two sites.  For the histograms in Figure \ref{fig:projected-correlator}, 32768 independent rounds of sweeps and measurements were used, and one plot was made for each vectorial argument of the correlator $r_i-r_j$. Measurements were spaced one Monte Carlo sweep apart, enabling a check for potential sampling of different symmetry breaking orders, while preserving measurement independence since one sweep is also approximately the autocorrelation time of the algorithm. In a finite system with $L^2$ unit cells, the value of a single measurement of each correlator $C(r)$ is quantized in units of $1/L^2$, which provided the natural bin size for the histogram.

\optionaltext{
\section{Algebraic decay of density-density correlator in the superfluid}
The correlations of density fluctuations above the mean, \optionaltext{Eq.~\ref{eq:densitycorr}, }can in principle distinguish a superfluid from a Mott insulator, as we shall now describe. 
The bosonic system has a conserved U(1) charge. If the boson $\hat{n}$ condenses, its conjugate operator $\theta$ gains an expectation value. Then\optionaltext{, just like the number current gains a contribution proportional to $\nabla \theta$,} the boson density gains a contribution
\beq
n-n_0 = \chi \partial_t \theta
\eeq
with $\chi$ the compressibility.
Evaluating the correlation function
\beq
\left< \partial_t \theta(r) \partial_t \theta(0) \right>
\eeq
at equal time and large spatial separation $r$ yields the algebraic form $\sim 1/r^3$. 

Now observe that if a more complicated combination of boson operators condenses, the operator $\theta$ conjugate to this combination gains an expectation value and is still described by a simple XY model. Moreover since there is just one global U(1) charge, the original boson density $n-n_0$ must still be proportional to $\partial_t \theta$ with some proportionality constant. \optionaltext{We conclude that whenever the correlator Eq.~\ref{eq:densitycorr} unambiguously exhibits exponential rather than algebraic decay at large $r$, the state is not a superfluid. }However, since for arbitrary condensation the proportionality constant relating $n$ to $\partial_t \theta$  is undetermined, it is difficult to rule out with numerical data any potential algebraic decay at large $r$ with a small prefactor. For the state $|P\Psi_{\hexagon}\rangle$, the rapid decay of density correlations supports the expectation that suppressing charge fluctuations by the projection $P$ would not turn $|\Psi_{\hexagon}\rangle$ into a superfluid.  
}

\end{appendix}

\bibliography{mottbib}

\end{document}